\documentclass[aps,prl,twocolumn,showpacs]{revtex4}
\usepackage{graphicx,amssymb,amsfonts,amsmath}

\begin{document}

\title{Nonequilibrium dynamics of a two-channel Kondo system due to a quantum quench
}
\author{Zurab Ratiani and Aditi Mitra}
\affiliation{Department of Physics, New York University, 4
Washington Place, New York, New York 10003 USA
}
\date{\today}


\begin{abstract}
Recent experiments by Potok et al. have demonstrated a remarkable tunability between
a single-channel Fermi liquid fixed point and a two-channel non-Fermi liquid fixed point.
Motivated by this we study the nonequilibrium dynamics due to a sudden quench of
the parameters of a Hamiltonian from
a single-channel to a two-channel anisotropic
Kondo system. We find a distinct difference between the long time behavior of
local quantities related to the impurity spin as compared to that of  bulk quantities
related to the total (conduction electrons + impurity) spin of the system.
In particular, the
local impurity spin and the
local spin susceptibility are found to equilibrate, but in a very slow power-law fashion
which is peculiar to the
non-Fermi liquid properties of the Hamiltonian.  In contrast, we find a
lack of equilibration in the two particle expectation values related to the total spin
of the system.
\end{abstract}

\pacs{71.27.+a, 72.10.Fk, 5.70.Ln}

\maketitle

The behavior of a local spin coupled to one or more independent
channels of conduction electrons is a classic problem in condensed matter physics. 
It is well known that when the
local spin has a size $S=1/2$ and is coupled to only
a single channel of conduction electrons, the spin is completely screened and 
the many particle
system behaves as a Fermi liquid~\cite{Hewson}.
In contrast, when the local spin is
coupled to two or more screening channels, one has dramatically different 
behavior where the spin is overscreened, and
the system exhibits non-Fermi liquid properties~\cite{Nozieres2CK}.
Due to
the success in realizing nanostructures consisting of a spin coupled to
one or more reservoirs, there has been a resurgence of interest in these classic systems. 
The primary focus
now is on understanding their nonequilibrium properties, such as the effect of current 
flow~\cite{Neqscalingexps}
and their nonequilibrium time evolution~\cite{timev}.

In this paper we will study nonequilibrium dynamics of a two-channel Kondo system. We are motivated by
recent experiments by Potok et al.~\cite{Potok07} where by tuning external gate voltages a local
spin could be effectively coupled to a single screening channel or 
to two independent screening channels. 
Thus as a function of gate voltage,
both single-channel Fermi liquid physics 
as well as two channel non-Fermi liquid physics was demonstrated 
on the same device.
We will study what happens when this external gate voltage
controlling the fine tuning is changed rapidly in time from an initial value corresponding to a 
single channel Kondo (1CK) system to a final value corresponding to
a two channel Kondo (2CK) system, thus inducing nonequilibrium dynamics. The time evolution of both 
single particle expectation values, as well as two particle
expectation values that exhibit non-Fermi liquid behavior in equilibrium will be studied.

Our system consists of two chiral
non-interacting fermions that constitute the two channels that interact with the local spin
$S =1/2$. We will employ the Emery-Kivelson mapping onto an interacting 
resonant level model~\cite{Kivelson92}. 
In the past any dynamics using this mapping, besides addressing other physical situations, 
was studied only at the non-interacting Toulouse point~\cite{Komnik09}. 
In this paper, in order to capture any nontrivial dynamics of the total spin of the system, we 
will have to move away from the Toulouse point. 
The Hamiltonian $H$ is: 
\begin{eqnarray}
&&
H = i v_F \sum_{\alpha,i=1,2} \int_{-\infty}^{\infty} dx \, {\psi}^{\dagger}_{i\alpha}
\frac{\partial}{\partial x}\psi_{i\alpha}(x) + \frac{h_1}{2} \tau^z  \nonumber\\
&&
+ \frac{h_2}{2} \sum_{i=1,2}\int dx  \left(\psi^{\dagger}_{i\uparrow}
\psi_{i\uparrow} - \psi^{\dagger}_{i\downarrow}\psi_{i\downarrow}\right)
\nonumber \\&&
+ \frac{1}{2}\sum_{\substack{\alpha\beta,i=1,2\\\lambda=x,y,z}}
J^{\lambda}_{i}(t) \tau^{\lambda}  {\psi}^{\dagger}_{i\alpha}(0) 
\sigma^{\lambda}_{\alpha \beta}\psi_{i\beta}(0)
\label{H0}
\end{eqnarray}
Above, $i$ labels
channel while  $\alpha,\beta$ labels spin index. $\vec{\tau},\vec{\sigma}$
are Pauli matrices and $\frac{1}{2} \psi^{\dagger}_{i\alpha}\vec{\sigma}_{\alpha \beta}
\psi_{i\beta}$ is the spin density operator for the electrons in the $i$-th channel, while
$\vec{S}= \vec{\tau}/2$ is the impurity spin operator. The coupling to the leads
$J^{\lambda}_{i}(t)$ are time-dependent (in an experiment these may be tuned by external gate voltages)
~\cite{Potok07}.
When $h_1=h_2$, a uniform magnetic field couples equally to both the impurity
spin as well as the spins in the leads. When $h_2=0, h_1 \neq 0$, the magnetic field couples
only to the impurity spin. We will assume anisotropic couplings $J^x_i(t)=J^y_i(t)=J^{\perp}_i(t)
\neq J^z_i(t)$.
It is convenient to define,
$\bar{J}_{z,\perp} = \left(J^{z,\perp}_1 + J^{z,\perp}_2\right)/2,
\delta J_{z,\perp} = J^{z,\perp}_1 - J^{z,\perp}_2$,
where $\delta J_{z,\perp}=0$ at the 2CK fixed point.

We briefly review the steps
involved in mapping the above model onto an interacting resonant level model~\cite{Kivelson92}.
One defines the canonically conjugate variables
$\left[\phi_{i\alpha}(x), \Pi_{j\beta}(y)\right] = i \delta_{ij} \delta_{\alpha \beta} \delta(x-y)$
in terms of which the fermions are written as
$\psi_{i\alpha}(x) = \exp{\left(-i \Phi_{i\alpha}(x)\right)}\eta_{i\alpha}/\sqrt{2\pi \alpha}$
where
$\Phi_{i\alpha}(x) = \sqrt{\pi} \left[\int_{-\infty}^x dx^{\prime} \Pi_{i\alpha}(x^{\prime})
+ \phi_{i\alpha}(x)\right]$.
This ensures that the same species of fermions anti-commute with each other.
$\eta_{i\alpha}$ are the Klein factors that are necessary to ensure
anti-commutation between different species of fermions. We choose~\cite{Schofield97}
$\eta_{i\alpha} = \exp{(i \theta^K_{i\alpha})}$ where,
$\theta^K_{1\downarrow} = 0; \theta^K_{1\uparrow} = \pi N_{1\downarrow};
\theta^K_{2\downarrow} = \pi\left(N_{1\downarrow} + N_{1\uparrow}\right);
\theta^K_{2\uparrow} = \pi \left(N_{1\downarrow} + N_{1\uparrow} + N_{2\downarrow}\right)$,
$N_{i\alpha}$ being the total number of $i\alpha$ fermions.
Defining $\chi_{i\alpha} = \Phi_{i\alpha} -\theta^K_{i\alpha}$,  
one changes variables to
$2\chi_c = \chi_{1\uparrow} + \chi_{1\downarrow}+\chi_{2\uparrow}+\chi_{2\downarrow}$,
$2\chi_s = \chi_{1\uparrow} - \chi_{1\downarrow}+\chi_{2\uparrow}-\chi_{2\downarrow}$,
$2\chi_f= \chi_{1\uparrow} + \chi_{1\downarrow} - \chi_{2\uparrow}- \chi_{2\downarrow}$,
$2\chi_{sf} =\chi_{1\uparrow} - \chi_{1\downarrow} -\chi_{2\uparrow}+ \chi_{2\downarrow}$.
Next one performs the unitary transformation
$H \rightarrow U^{\dagger}HU$, where
$U = \exp\left[-i S_z \chi_s(0)\right]$, followed by a refermionization of the
Hamiltonian into the fermionic fields  
$d^{\dagger}=-i S^+ ;\, d = i S^{-}$ (so that $d^{\dagger} d - \frac{1}{2} = S^z$)
and $\psi_{\nu=c,s,f,sf}(x) = e^{i\pi d^{\dagger}d} e^{-i\chi_{\nu}(x)}/\sqrt{2\pi \alpha}$. 
In what follows we will assume
that $J^z_{1,2}, J_{1\perp}$ are
time independent and $\delta J^z = 0$. We set $J_{1\perp}=J_{\perp}$, and the only time-dependence will
be in $J_{2\perp}(t)$. With this we obtain,
\begin{eqnarray}
&&U^{\dagger} HU = iv_F\sum_{\nu=c,s,f,sf} \int_{-\infty}^{\infty}
dx \left(\psi_{\nu}^{\dagger}
\frac{\partial \psi_{\nu}}{\partial x}\right) \label{H1e}\\
&&+ h_2\int_{-\infty}^{\infty} dx
\psi^{\dagger}_s(x) \psi_s(x)
+\left(h_1 - h_2\right)\left(d^{\dagger}d -\frac{1}{2}\right) \label{h}\\
&&+ 2\left(\bar{J}^z - \pi v_F\right) \left(d^{\dagger}d -\frac{1}{2}\right)
:\psi_{s}^{\dagger}(0)\psi_{s}(0): \\
&&
+ \frac{{J}_{\perp}}{\sqrt{2\pi\alpha}}\left[d^{\dagger}
\psi_{sf}(0) + \psi^{\dagger}_{sf}(0) d  \right]   \\
&&+ \frac{J_{2 \perp}(t)}{\sqrt{2\pi \alpha}}\left[d^{\dagger}
\psi^{\dagger}_{sf}(0) + \psi_{sf}(0) d  \right] \label{H1e1}
\end{eqnarray}
We will assume that the time dependence of 
$J_{2\perp}(t)$ is that of a quench, $J_{2\perp}(t) = J_{\perp}\theta(t)$.
Thus for $t<0$ we have a 1CK system that is described by an
interacting resonant level model. Whereas for $t>0$, the Hamiltonian is that of a 2CK
system where the coupling of the resonant level to the reservoir of $\psi_{sf}$ fermions is 
via $\frac{{J}_{\perp}}{\sqrt{2\pi\alpha}}\left[(d^{\dagger}-d)
\psi_{sf}(0) + h.c. \right] $. Thus in the 2CK model effectively only half of the resonant 
level corresponding to
the Majorana fermion $a=-i(d^{\dagger}-d)/\sqrt{2}$
couples to the conduction electrons, while the other half $b=(d^{\dagger}+d)/\sqrt{2}$, does not couple.
As was pointed out in~\cite{Kivelson92} all non-Fermi liquid behavior
stems from this peculiarity of the resonant level, and as we shall see is also responsible 
for interesting behavior in
the dynamics.

To see this note that immediately after the quench
we have a highly nonequilibrium system, where any local degrees of freedom can relax
to the ground state only via their coupling to the reservoirs. In the 2CK model since only half the
local degrees of freedom are coupled, local quantities relax very slowly, as we shall show
in a power law manner. Moreover we find that two particle expectation values related to the total 
(bulk + local) spin of the system do not relax to their equilibrium values.

{\bf Time evolution of local quantities:}
We will first consider the case when the external magnetic field couples only to the local
spin, so that $h_1=h$ and $h_2=0$. 
We will study the time evolution
of the local magnetization, and the local spin susceptibility, the latter in the limit
$h \rightarrow 0$.
To capture the non-Fermi liquid behavior of the
local susceptibility in an equilibrium 2CK system, it suffices to be at the
noninteracting Toulouse point
$\bar{J}^z=\pi v_F$. Therefore the nonequilibrium dynamics of the local 
quantities will also be studied at
the Toulouse point. Later while studying the dynamics of the total spin, we will have to move 
away from the 
Toulouse point so as 
to capture non-Fermi liquid physics~\cite{Clarke93,Georges94}.
We define the following Green's functions for the local fermion (spin),
\begin{eqnarray}
\hat{G}^R(t,t^{\prime}) = -i \theta(t-t^{\prime}) \langle\{\begin{pmatrix} d(t) \\ d^{\dagger}(t)
\end{pmatrix},
\begin{pmatrix} d^{\dagger}(t^{\prime})& d(t^{\prime})\end{pmatrix}\} \rangle \label{GRdef}\\
\hat{G}^K(t,t^{\prime}) = -i\langle\left[\begin{pmatrix} d(t) \\ d^{\dagger}(t)
\end{pmatrix},
\begin{pmatrix} d^{\dagger}(t^{\prime})& d(t^{\prime})\end{pmatrix}\right] \rangle \label{GKdef}
\end{eqnarray}
Denoting the individual elements of the above matrices as
$\hat{G} =\begin{pmatrix} G_{d,d^{\dagger}} & G_{d,d} \\ G_{d^{\dagger},d^{\dagger}}
& G_{d^{\dagger},d}
\end{pmatrix} $, $G^{R,K}$ obey the equation of motion:
\begin{eqnarray}
&&\left[i\partial_t  -h \begin{pmatrix} 1&0\\ 0 & -1\end{pmatrix} -\hat{\Sigma}^R
\circ\right] \hat{G}^R =1 \label{GRtim} \\
&&\hat{G}^K = \hat{G}^R \circ \hat{\Sigma}^K \circ \hat{G}^A \label{GKtim}
\end{eqnarray}
where $\circ$ denotes convolution in time,
$G^A(t,t^{\prime}) = \left[ G^R(t^{\prime},t)\right]^*$, and
$\Sigma^{R,K}$ are the self-energies due to
coupling to the leads. Defining $\Gamma_{\perp} = \frac{J_{\perp}^2}{\pi \alpha v_F}$,
a time-dependence of the form $J_{2\perp}(t) = J_{\perp}\theta(t)$ implies,
\begin{eqnarray}
&&\hat{\Sigma}^R(t,t^{\prime}) = \frac{-i\Gamma_{\perp}}{4} \delta(t-t^{\prime})\begin{pmatrix}
1 + \theta^2(t) & -2\theta(t) \\ -2\theta(t)
& 1 + \theta^2(t)\end{pmatrix}\label{SigmaR}\\
&&\hat{\Sigma}^K(t,t^{\prime}) = 
-\frac{\Gamma_{\perp}}{2}P\left(\frac{T}{\sinh{\pi T (t-t^{\prime})}} 
\right) \times \label{SigmaK}\\
&&\times \begin{pmatrix}
1 + \theta(t) \theta(t^{\prime})
& -\left(\theta(t^{\prime}) + \theta(t)\right)
\\
-\left(\theta(t^{\prime}) + \theta(t)\right) &
1 + \theta(t) \theta(t^{\prime})
\end{pmatrix} \nonumber 
\end{eqnarray}
where $T$ is the temperature of the conduction electrons.

The solutions to Eq.~(\ref{GRtim}) 
depend on whether the time arguments 
in $G^R(t,t^{\prime})$
are before or after the quench.
When both times are before the quench,
\begin{eqnarray}
&&
\hat{G}^R(t<0,t^{\prime}<0) = 
\label{GRsol1a} \\
&&
-i\theta(t-t^{\prime})e^{-\frac{\Gamma_{\perp}}{4}\left(t-t^{\prime}\right)}
\begin{pmatrix}
e^{-ih\left(t-t^{\prime}\right)}&0
\\0 &e^{ih\left(t-t^{\prime}\right)}
\end{pmatrix} \nonumber
\end{eqnarray}
When both times are after the quench we get, 
\begin{eqnarray}
&&\hat{G}^R(t>0,t^{\prime}>0) = -i\theta(t-t^{\prime})
e^{-\frac{\Gamma_{\perp}}{2}\left(t-t^{\prime}\right)}
\nonumber\\
&&\begin{pmatrix} A_1(t,t^{\prime})& B_1(t,t^{\prime}) \\
B_1(t,t^{\prime}) & \left[A_1(t,t^{\prime})\right]^*
\end{pmatrix}\label{GRsol1b}
\end{eqnarray}
where 
$A_1(t,t^{\prime})=\!\!\!\!\cosh{\left(\sqrt{\frac{\Gamma_{\perp}^2}{4}-h^2}(t-t^{\prime})\right)}
-\frac{i h}{\sqrt{\frac{\Gamma_{\perp}^2}{4}-h^2}}
\sinh{\left(\sqrt{\frac{\Gamma_{\perp}^2}{4}-h^2}(t-t^{\prime})\right)}$,
$B_1(t,t^{\prime})=\frac{\Gamma_{\perp}/2}{\sqrt{\frac{\Gamma_{\perp}^2}{4}-h^2}}
\sinh{\left(\sqrt{\frac{\Gamma_{\perp}^2}{4}-h^2}(t-t^{\prime})\right)}$.
When one of the times is after the quench and the other before,
\begin{eqnarray}
&&\hat{G}^R(t>0,t^{\prime}<0) =
 -i\theta(t-t^{\prime})e^{-\frac{\Gamma_{\perp}t}{2} +
\frac{\Gamma_{\perp}t^{\prime}}{4}}
\nonumber\\
&&\begin{pmatrix} A_2(t,t^{\prime})& B_2(t,t^{\prime}) \\
B_2(t,t^{\prime}) & \left[A_2(t,t^{\prime})\right]^*
\end{pmatrix} \label{GRsol1c}
\end{eqnarray}
where $A_2(t,t^{\prime})= \cosh{\sqrt{\left(\frac{\Gamma_{\perp}t}{2}\right)^2 - h^2 (t-t^{\prime})^2}}
-\frac{i h(t-t^{\prime})}{\sqrt{\left(\frac{\Gamma_{\perp}t}{2}\right)^2 - h^2 (t-t^{\prime})^2}}
\sinh{\sqrt{\left(\frac{\Gamma_{\perp}t}{2}\right)^2 - h^2 (t-t^{\prime})^2}}$
,$B_2(t,t^{\prime})= \frac{\Gamma_{\perp}t/2}{\sqrt{\left(\frac{\Gamma_{\perp}t}{2}\right)^2
- h^2 (t-t^{\prime})^2}}
\sinh{\sqrt{\left(\frac{\Gamma_{\perp}t}{2}\right)^2 - h^2 (t-t^{\prime})^2}}$.

We first discuss the behavior of the magnetization $S^z = \frac{-i}{2}G^K_{d,d^{\dagger}}(t,t)$ for
a time after the quench (hence, $t >0$) and when the conduction electrons are at a temperature $T=0$.
Substituting Eq.~(\ref{SigmaK}),~(\ref{GRsol1b}),~(\ref{GRsol1c}) in Eq.~(\ref{GKtim}), 
in the limit of
$h \ll \Gamma_{\perp}$ and long times $t \gg 1/h$ we find,
\begin{eqnarray}
S^z(t) - S^z_{2CK,eq} \sim  e^{-\frac{h^2}{\Gamma_{\perp}}t}
\left[\frac{1}{\pi h t} + {\cal O}\left(\frac{1}{h^2 t^2}, \frac{1}{\Gamma_{\perp}t}\right) \right]
\label{Szt}
\end{eqnarray}
where $ S^z_{2CK,eq}= -\frac{h}{2\pi \sqrt{\Gamma^2_{\perp}-4 h^2}}
\ln\left[\frac{\Gamma_{\perp}^2 - 2 h^2 +\Gamma_{\perp}\sqrt{\Gamma_{\perp}^2 - 4 h^2}}
{\Gamma_{\perp}^2 - 2 h^2 -\Gamma_{\perp}\sqrt{\Gamma_{\perp}^2 - 4 h^2}}\right]$ is
the local magnetization in the ground state of the 2CK Hamiltonian. 
Thus the local magnetization
does equilibrate, but at a slow rate of $h^2/\Gamma_{\perp}$ associated with the
$b$ fermion. In contrast, for a reverse
quench $J_{2\perp}(t)=J_{\perp}\theta(-t)$ where the time evolution is governed by a 1CK model
for which the $a$ and $b$ fermions are equally coupled to the reservoirs, 
we have checked that $S_z$ equilibrates at the much faster rate of $\Gamma_{\perp}/4$.

We will now study the time evolution of 
the local longitudinal spin response
function 
$\chi^R_{loc}(t,t^{\prime}) = -i \theta(t-t^{\prime})
\langle \{d^{\dagger}(t) d(t), d^{\dagger}(t^{\prime})
d(t^{\prime})\} \rangle$ which we rewrite as,
\begin{eqnarray}
&&
\chi^R_{loc}(t,t^{\prime})= \nonumber \\
&&\frac{-i}{2} \left[G^R_{d,d^{\dagger}}(t,t^{\prime}) G^K_{d,d^{\dagger}}(t^{\prime},t)
 + G^K_{d,d^{\dagger}}(t,t^{\prime}) G^A_{d,d^{\dagger}}(t^{\prime},t) \right. \nonumber \\
&&\left. - G^R_{d^{\dagger},d^{\dagger}}(t,t^{\prime}) G^K_{d,d}(t^{\prime},t)
-G^K_{d^{\dagger},d^{\dagger}}(t,t^{\prime}) G^A_{d,d}(t^{\prime},t)\right]
\label{chiRloc1}
\end{eqnarray}
It is useful to define the nonequilibrium static susceptibility at time $T_m$,
$\chi_{S,loc}(T_m) = \int_0^{\infty} d\tau \chi^R_{loc}(T_m + \frac{\tau}{2},T_m-\frac{\tau}{2})$.
For $h=0$, and for very low temperatures $T \ll \Gamma_{\perp}$
of the conduction electrons,
we find the following behavior for the static susceptibility at
times $T_m \gg 1/\Gamma_{\perp}$,
\begin{eqnarray}
&&\chi_{S,loc}(T_m) -\chi^{eq}_{S,loc,2CK} \label{locte} \\
&&\sim \frac{1}{\pi \Gamma_{\perp}}\ln\left(\frac{1}{2T T_m}\right)
+ \frac{1}{\pi \Gamma_{\perp}}{\cal O}\left(\frac{1}{\Gamma_{\perp} T_m}\right)
\forall   T T_m \ll 1 \nonumber \\
&& \sim \frac{1}{\pi \Gamma_{\perp}} \left(\frac{1}{2 T T_m}\right)
\forall   T T_m \gg 1 \nonumber
\end{eqnarray}
where~\cite{Kivelson92,Georges94}
$\chi^{eq}_{S,loc,2CK} = \frac{1}{\pi \Gamma_{\perp}}
\ln{\frac{T}{\Gamma_{\perp}}}$
is the equilibrium (non-Fermi liquid)
local susceptibility of the 2CK Kondo system. 
Thus we find that the logarithmic singularity associated with the 2CK system is
cut-off by $max\left(T,\frac{1}{T_m}\right)$. Moreover at long times $TT_m\gg 1$, 
the local susceptibility equilibrates, but in a very slow power law fashion which is determined
by the temperature of the leads.

{\bf Time evolution of bulk + local quantities :}
Let us consider the case where an external magnetic field couples to the total
(conduction electrons + local) spin of the system so that
$h_1=h_2=h$. 
We will discuss the time evolution of the
response function of the total spin of the system when $h\rightarrow 0$.
From Eq.~\ref{h}, this may be formally defined as
$\chi^R(x,t;yt^{\prime}) =
-i\theta(t-t^{\prime})
\langle \{\psi_s^{\dagger}(x,t) \psi_s(x,t), \psi^{\dagger}_s(y,t^{\prime})\psi_s(y,t^{\prime})\} 
\rangle $.
At the Toulouse point $\bar{J}^z=\pi v_F$, the local
degrees of freedom do not couple to the bulk field $\psi_s$, so that the response
function is independent of the local quench and
is given by the Lindhard function,
\begin{eqnarray}
\chi^R_0(q,\Omega) 
&&=
\left(\frac{-L}{2\pi v_F}\right) \frac{q v_F}{q v_F -\left(\Omega + i \delta\right)}
\label{lind1}
\end{eqnarray}
Thus the static spin susceptibility
$\chi_{S0}(q,\Omega=0) = \left(\frac{-L}{2\pi v_F}\right)$, and
is independent of $q$.

\begin{figure}
\includegraphics[totalheight=1.5cm,width=5cm]{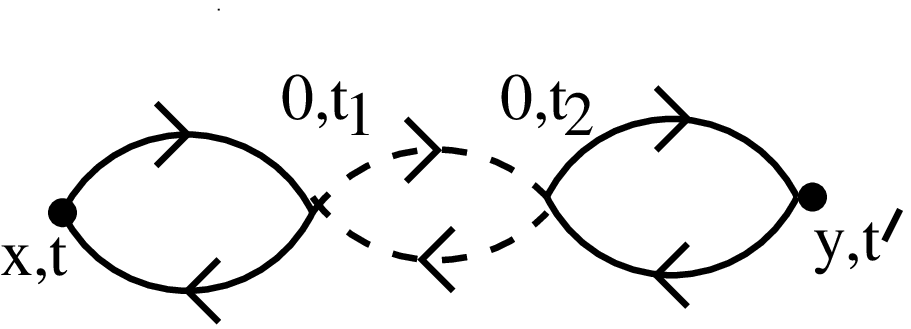}
\caption{Diagram for $\chi^R_{imp}(xt,yt^{\prime})$. Dark line: 
propagator for the conduction electrons $\psi_s$. Dashed line: propagator for the
local $d$ fermion.}
\label{diag}
\end{figure}

To obtain non-Fermi liquid behavior one has to move away from the Toulouse 
point~\cite{Clarke93,Georges94},
which couples $\psi_s$ to the local field, and also introduces
non-equilibrium dynamics in $\chi^R$.
Defining,
$\chi^R(q; t, t^{\prime}) = \int dx \int dy \cos{q(x-y)} \chi^R(x,t;yt^{\prime})$),
the leading correction in $\left(\bar{J}^z-\pi v_F\right)$ to $\chi^R$ (shown
in Fig.~\ref{diag}) is,
\begin{eqnarray}
&&\chi^R_{imp}(q;t,t^{\prime})= \nonumber \\
&&\left(\bar{J}^z-\pi v_F\right)^2\!\!
\int_{-\infty}^{\infty} \!\!dt_1 \int_{-\infty}^{\infty}dt_2 dx dy \,\,\,\,\chi^R_{loc}(t_1,t_2) \nonumber \\
&&\cos{(q(x-y))}
\left[G^R_{\psi_s^{\dagger}\psi_s}(x,0;t,t_1)
G^K_{\psi_s^{\dagger}\psi_s}(0,x;t_1,t) \right. \nonumber \\
&&\left.+ G^K_{\psi_s^{\dagger}\psi_s}(x,0;t,t_1) G^A_{\psi_s^{\dagger}\psi_s}(0,x;t_1,t)\right]
\nonumber \\
&&\times\left[G^R_{\psi_s^{\dagger}\psi_s}(0,y;t_2,t^{\prime})
G^K_{\psi_s^{\dagger}\psi_s}(y,0;t^{\prime},t_2) \right. \nonumber \\
&&\left. + G^K_{\psi_s^{\dagger}\psi_s}(0,y;t_2,t^{\prime})
G^A_{\psi_s^{\dagger}\psi_s}(y,0;t^{\prime},t_2)\right]
\end{eqnarray}
where we have assumed that the interaction $\left(\bar{J}^z-\pi v_F\right)$
has been switched on adiabatically slowly at long times in the past. 
The label $\chi_{imp}$ signifies that it is the correction to the bulk response-function
due to coupling to the local impurity,
$G^{R,K}_{\psi_s^{\dagger}\psi_s}$ are the Green's functions of the free $\psi_s$
fermions, and $\chi^R_{loc}$ is defined in Eq.~\ref{chiRloc1}.
Defining $t= T_m + \frac{\tau}{2}; t^{\prime} =  T_m - \frac{\tau}{2} $, the
nonequilibrium static
susceptibility $\chi_{S,imp}(q,T_m) 
= \int_0^{\infty} d\tau \chi^R_{imp}(q;T_m + \frac{\tau}{2},T_m - \frac{\tau}{2})$
is
\begin{eqnarray}
&&
\chi_{S,imp}(q,T_m) = \frac{1}{4}\left(\bar{J}^z-\pi v_F\right)^2\label{chiimpA} \\
&&\int dt_1 \int dt_2 \chi^R_{loc}(t_1,t_2)\int\frac{d\epsilon}{2\pi}
e^{-2i T_m \epsilon + i\epsilon(t_1+t_2)}
\nonumber \\
&& \frac{1}{L^2}
\left[\chi_0^R(q,\epsilon)\chi_0^R(q,-\epsilon)+ 
\chi_0^R(-q,\epsilon)\chi_0^R(-q,-\epsilon)\right]\nonumber
\end{eqnarray}
where $\chi^R_0$ is given in Eq.~(\ref{lind1}).
In equilibrium i.e., when $J_2 = J_{\perp}$ and is independent of time,
$\chi^R_{loc}$ is independent of $t_1 + t_2$. Thus the time integral over
$t_1+t_2$ forces $\epsilon=0$ in Eq.~(\ref{chiimpA}). With this one
recovers the equilibrium result~\cite{Clarke93,Georges94}
$\chi_{S,imp}(q)
=  \frac{1}{4}\left(\frac{\bar{J}^z-\pi v_F}{2\pi v_F}\right)^2
\chi^{eq}_{S,loc,2CK}=\chi^{eq}_{S,imp,2CK}$.
To study the evolution of the static susceptibility after the quench, it is
convenient to change variables in Eq.~(\ref{chiimpA}) to $T^{\prime} =\frac{t_1+t_2}{2},
\tau = t_1 - t_2$. Defining $u=\left(\frac{\bar{J}_z-\pi v_F}{2\pi v_F}\right)$
and performing the integration over $\epsilon$ we get,
\begin{eqnarray}
&&\chi^R_{S,imp}(q,T_m) =
\frac{u^2}{4}
\int dT^{\prime} \int d\tau
\nonumber \\
&&\chi^R_{loc}(T^{\prime} + \frac{\tau}{2},T^{\prime} - \frac{\tau}{2})
\left[q v_F\sin{(2 q v_F|T^{\prime}-T_m|)}\right]
\label{chiimpC}
\end{eqnarray}
We will present results for $q v_F \ll \Gamma_{\perp}$ and
times $T_m \gg 1/\Gamma_{\perp}$ so that terms that fall off as
$\frac{1}{\Gamma_{\perp} T_m}$ or faster will be dropped. Further, we 
will consider two cases: one where $q =0$, and the other when 
$qv_F \gg (T,\frac{1}{T_m})$.

For $q=0$, note that we should first perform the $T^{\prime}$ integral in Eq.~(\ref{chiimpC}), and then
set $q=0$. This gives,
\begin{eqnarray}
&&{\chi}_{S,imp}(q=0) = \frac{1}{2}\chi^{eq}_{S,imp,2CK} 
+\frac{1}{2}\chi^{eq}_{S,imp,1CK} 
\label{ansq0}
\end{eqnarray}
where $\chi^{eq}_{S,imp,1CK}=-\left(\frac{\bar{J}_z-\pi v_F}
{2 \pi v_F}\right)^2\frac{1}{\pi \Gamma_{\perp}} $
is the static susceptibility in the 1CK ground state.
For the case $qv_F \gg \left(T,\frac{1}{T_m}\right)$, dropping terms of
${\cal O}\left(\frac{1}{2 q v_F T_m}\right)$, we find 
\begin{eqnarray}
&&\chi_{S,imp}(qv_F) 
= \chi^{eq}_{S,imp,2CK}\left(1 - \frac{1}{2}\cos{(2qv_FT_m)} \right)
\nonumber  \\
&&+ \frac{1}{2}\chi^{eq}_{S,imp,1CK}\cos{(2qv_FT_m)}
- \frac{u^2}{4\pi \Gamma_{\perp}} CosInt\left(4 T T_m\right)\nonumber \\
&&-\frac{u^2}{8\pi \Gamma_{\perp}}\left[\frac{q^2v_F^2}{\Gamma_{\perp}^2} 
\ln\frac{\Gamma_{\perp}}{2T} + \left(\ln\frac{q v_F}{2T}\right) + \ldots \right]
\cos{(2qv_FT_m)}\nonumber \\
&&-\frac{u^2}{20\pi \Gamma_{\perp}}\frac{q^2 v_F^2}{\left(\Gamma_{\perp}^2 
+ q^2 v_F^2\right)}g(TT_m)
\label{van1}
\end{eqnarray}
where $g(x\ll 1) \sim 1 + {\cal O}(x^2)$, $g(x\gg 1)\sim \frac{1}{x}$, and
$\ldots$ represent terms that are small in comparison to $\ln\left(\frac{qv_F}{2T}\right), 
\ln\left(\frac{\Gamma_{\perp}}{2T}\right)$.

Thus we find a marked difference between the susceptibility at long times 
after the quench and the 
susceptibility in equilibrium $\chi^{eq}_{S,imp,2CK}$. While $\chi^{eq}_{S,imp,2CK}$
is independent of wave-vector, the out of equilibrium susceptibility is strongly dependent on $q$,
and does not even reach a time independent steady state, but instead  
oscillates at frequency $q v_F$ (Eq.~(\ref{van1})). 
For intermediate times $TT_m\ll 1$, performing a time-averaging so that terms that oscillate at
$qv_F$ go to zero, we find,
\begin{eqnarray}
&&\bar{\chi}_{S,imp}(qv_F \gg \frac{1}{T_m} \gg T) = \frac{u^2}{4\pi \Gamma_{\perp}}\left[ 
\ln\frac{1}{4 \Gamma_{\perp} T_m} \right. \nonumber \\
&&\left. - \frac{q^2 v_F^2}{5\left(\Gamma_{\perp}^2 
+ q^2 v_F^2\right)} \right]
\label{ansqneq0}
\end{eqnarray}

Thus for an intermediate time which is longer, the lower the temperature, the logarithmic
divergences associated with the bulk susceptibility 
$\bar{\chi}_{S,imp}(qv_F)$ not only get cutoff by inverse-time (a result similar to
Eq.~\ref{locte} for the local susceptibility), it also acquires some $q$-dependent 
corrections.
In contrast, at long times $TT_m\gg1$, Eq.~(\ref{van1}) implies that the
time-averaged susceptibility at large wave-vectors
$qv_F \gg T$ is,
$\bar{\chi}_{S,imp}(qv_F \gg T \gg \frac{1}{T_m}) = \chi^{eq}_{S,imp,2CK}  
+ {\cal O}\left(\frac{1}{T T_m}\right)$, 
and therefore equilibrates. 

The $q=0$ static susceptibility (Eq.~(\ref{ansq0})) on the other hand is found to reach a time
independent steady state which is an equal mixture of the non-analytic
in temperature form of the 2CK ground state, and the analytic in 
temperature form of the 1CK ground state.
This lack of equilibration in bulk properties is consistent with nonequilibrium time evolution in 
integrable models where the system retains memory of its initial state. For local quantities on 
the other hand (Eq.~(\ref{Szt}),~(\ref{locte})), at least at the
Toulouse point, the rest of the system to which they are coupled 
acts as a reservoir causing them to equilibrate, but at very slow rates compared to a 1CK model. 

In summary we have studied the nonequilibrium dynamics in a 2CK system due to a quantum quench.
Our results highlight how the non-Fermi liquid properties of the system, along with its integrability
affect the time evolution of single particle and two-particle expectation values. 
An interesting question concerns the observability of the nonequilibrium dynamics presented here.
Experiments may be characterized by two kinds of effects that have not been taken into account
in the present treatment. One is that the system could be ``open'' {i.e}, coupled to some other modes
such as phonons, leading to an external dissipation rate $\gamma_{diss}$ which will eventually cause the system
to equilibrate. The second effect could be deviations from integrability arising for example due to a nonlinear
dispersion for the conduction electrons. Studying the consequence of these effects is very interesting and
beyond the scope of this paper. However, one may still be able to speculate on the effect of an
external dissipation. 
In particular a characteristic
of the 2CK system is slow power-law dynamics. Thus we expect that for weak dissipation 
$\gamma_{diss}\ll \Gamma_{\perp}$, the system will equilibrate slowly as $\frac{1}{\gamma_{diss}T_m}$, 
(where $T_m$ is the time after the quench) so that a nonequilibrium/transient state can still exist for 
long enough time-scales to be observable. The results of this paper are also relevant for
Kondo systems in cold-atom gases where dissipative effects are weak~\cite{2CKcoldatoms}. 

{\it Acknowledgments:}
This work was supported by NSF-DMR (Contract No: 0705584).

\end{document}